\documentstyle[sprocl]{article}

\def\Journal#1#2#3#4{{#1} {\bf #2}, #3 (#4)}

\def\NPB{{\em Nucl. Phys.} B}
\def\PLB{{\em Phys. Lett.}  B}

\def\PRD{{\em Phys. Rev.} D}

\def\be{\begin{equation}}
\def\ee{\end{equation}}
\def\bea{\begin{eqnarray}}
\def\eea{\end{eqnarray}}

\begin{document}

\title{POSITIVITY CONSTRAINTS ON ANOMALIES AND SUPERSYMMETRY}

\author{A. JOHANSEN}

\address{Lyman Laboratory of Physics,  Harvard University, 
Cambridge,\\ MA 02438, USA\\E-mail: johansen@pauli.harvard.edu} 




\maketitle\abstracts{ The relation between the trace and R-current anomalies
in 4D supersymmetric theories implies that the U(1)$_R$F$^2$, U(1)$_R$
and U(1)$^3_R$ anomalies which matched in studies of N=1 Seiberg duality
satisfy positivity constraints.
These constraints are tested in a large number of N=1 supersymmetric 
gauge theories in the non-Abelian 
Coulomb phase, and they are satisfied in all renormalizable models with 
unique anomaly-free R-current, including those with accidental symmetry.
Most striking is the fact that the flow of the Euler anomaly coefficient,
$a_{UV}-a_{IR}$, is always positive, as conjectured by Cardy.}

\section{2D theories}

This talk\footnote{Talk given at PASCOS-98.} 
is based on our joint papers \cite{AFG,AEF} with D. Anselmi,  J. Erlich, D. Freedman and 
M. Grisaru.

Various positivity constraints on anomalies have been known 
for long time
in the context of two-dimensional quantum field theories (QFT).
In this section I briefly remind a few aspects of renomalization group 
(RG) flow in 2D theories.

Consider a 2D non-critical unitary 
QFT which flows from the ultraviolet (UV) to the infrared
(IR) fixed point.
Let $T=T_{zz}$ denote a component of the stress tensor, where $z=x_1+ix_2$.
For simplicity we assume that there is only one coupling constant $g$ in the theory.
At a finite distance $x$ we have 
$\langle T(x) T(0)\rangle = c(g(1/x))/2z^4.$
Here $g(1/x)$ is a running coupling constant at the scale $1/|x|$.
The central function $c(g(1/x))$ interpolates between the UV value,  $c_{UV}$
at $x\to 0$ and the IR value, $c_{IR}$ at $x\to \infty$.
It follows from the unitarity that $c(g(1/x)) >0$, and hence
$c_{UV}, ~c_{IR}>0$.

The same theory can be considered in an external gravitational background
with a metric $g_{\mu\nu}$ and a curvature $R$.
The presence of external gravitational field induces an external anomaly term 
in the operator equation for the trace anomaly
$T_{\mu}^{\mu}=\tilde{c}(g(m)) R/12 + ~ internal ~anomaly,$ 
where $m$ is an RG scale.
Here the $internal ~anomaly$ term denotes a part of the trace anomaly that 
depends on the quantum fields.
The function $\tilde{c}(g(m))$ flows to $c_{UV}$ at $m\to \infty$ (the ultraviolet) and
to $c_{IR}$ at $m\to 0$ (the infrared).

Zamolodchikov's c-theorem  \cite{zamolo}
states that there exists a function
$c(g,x)=c(g(1/x))+ ~correction~ terms$ which
 i) is monotone decreasing toward the IR,
 ii) is stationary at the fixed points,
 iii) coincides with the central charges at the fixed points.
The {\it correction terms} vanish at the critical points so that $c_{UV}-c_{IR}>0$.
Zamolodchikov's c-function essentially counts the number of physical 
degrees of freedom. The  number of degrees of freedom in the IR is less than in the UV,
i.e. the RG flow is irreversible.
It also follows from the Zamolodchikov c-theorem that the central charge
at a critical point does not depend on the {\it marginal} deformations.

A natural question is how much of the above facts generalize to four dimensional
theories.
In 1988 Cardy \cite{cardy} conjectured that the Euler anomaly coefficient in the 
trace of stress tensor in the presence of external gravitational field
$T_{\mu}^{\mu}= a(g(m)) \tilde{R}\tilde{R} + ...$
universally obeys $a_{UV}-a_{IR}>0$.
(We shall refer to the inequality
$a_{UV} -a_{IR} >0$ as the $a$-theorem.)
This conjecture has been tested in the perturbation theory \cite{CFL}.
Also, some evidence supporting this conjecture  
has been found  \cite{bast} in the SU$(N_c)$ series of
SUSY gauge theories with $N_f$ fundamental quark flavors
where the conjectured electric-magnetic duality
\cite{seiberg}
allows to reformulate a strongly coupled (confining)
theory in terms of the 
weakly coupled magnetic one.
The basic techniques for computing the flow of
central charges when there is an interacting IR fixed point were
developed
\cite{AFG} and applied to the conformal phase for
$3N_c/2 < N_f < 3N_c.$
This approach does not use the electric-magnetic duality conjecture.
It has been applied \cite{AEF} to a large set of supersymmetric theories
where new
non-perurbative evidence for Cardy's conjecture has 
been found.
The formulas of refs. [1,2] have been checked 
\cite{FO} by explicit 
perturbative computation.
Finally, Forte and Latorre presented a proof
\cite{forte} of Cardy's conjecture.
In this talk I focus on the approach of refs. [1,2]. 

\section{Anomalies in 4D theories}

The computation of chiral anomalies of the $R$-current and
conserved flavor currents is one of the important tools used to
determine the non-perturbative infrared behavior of the many
supersymmetric gauge theories analyzed during the last few years.
The anomaly coefficients are subject to rigorous positivity
constraints by virtue of their relation to two-point functions of
currents and stress tensors, and to other constraints conjectured
in connection with possible four-dimensional analogues of the
Zamolodchikov c-theorem. 
The two-point functions
have been considered \cite{gof97} as central functions whose 
ultraviolet
and infrared limits define central charges of super-conformal theories
at the endpoints of the renormalization group flow.  The
positivity conditions are reasonably well known from studies of
the trace anomaly for field theories in external backgrounds.  In
supersymmetric theories the trace anomaly of the stress tensor
and conservation anomaly of the $R$-current are closely related,
which leads \cite{AFG} to positivity constraints on chiral
anomalies.  

The theoretical basis for the analysis of anomalies in supersymmetric
theories comes
from a combination of three fairly conventional ideas, namely

$\bullet$ The close relation between the trace anomaly of a 
four-dimensional field theory with external sources 
for flavor currents
  and stress tensor and the two point
  correlators $\langle J_{\mu} (x) J_{\nu}(y) \rangle$ and
  $\langle T_{\mu\nu} (x) T_{\rho \sigma} (y) \rangle$ and their
  central charges.

$\bullet$ The close relation in a supersymmetric theory between the
  trace anomaly $\Theta = T^{\mu}_{\mu}$ and the anomalous
  divergence of the $R$-current $\partial_{\mu}R^{\mu}$.

$\bullet$ The fact that anomalies of the $R$-current can be
  calculated at an infrared superconformal fixed point using 't
  Hooft anomaly matching. This is the standard procedure, and
  one way to explain it is to use the all orders anomaly free
  $S$-current of Kogan, Shifman, and Vainshtein \cite{kogan}.

We consider a supersymmetric gauge theory containing chiral
superfields $\Phi^{\alpha}_{i}$ in irreducible representations
$R_i$ of the gauge group $G$. 
To simplify the discussion we assume
that the superpotential $W=0$, but the treatment can be
generalized to include non-vanishing superpotential.

We consider a conserved current $J_{\mu}(x)$ for a non-anomalous
flavor symmetry $F$ of the theory, and we add a source
$B_{\mu}(x)$ for the current, effectively considering a new theory
with an additional gauged U$(1)$ symmetry but without kinetic
terms for $B_{\mu}$. The source can be introduced as an external
gauge superfield $B(x, \theta, \bar{\theta})$ so supersymmetry
is preserved. We also couple the theory to an external
supergravity background, contained in a superfield $H^{a} (x,
\theta, \bar{\theta})$, but we discuss only the vierbein $e^a_\mu(x)$ and the
component $V_{\mu} (x)$ which is the source for the $R^{\mu}$
current of the gauge theory.

The trace anomaly of the theory then contains several terms
\begin{equation}
\Theta=\frac{1}{2g^3}\tilde{\beta}(g)
(F_{\mu\nu}^a)^2 + \frac{1}{32 \pi^2} \tilde{b} (g) B^2_{\mu\nu}
+\frac{\tilde{c}(g)}{16\pi^2}(W_{\mu\nu\rho\sigma})^2-\frac{a(g)}
{16\pi^2}(\tilde{R}_{\mu\nu\rho\sigma})^2 +
\frac{\tilde{c}(g)}{6\pi^2}V_{\mu\nu}^2\, , \label{2.1}
\end{equation}
where $W_{\mu\nu\rho\sigma}$ is the Weyl tensor,
$\widetilde{R}_{\mu\nu\rho\sigma}$ is the dual of the curvature,
and $B_{\mu\nu}$ and $V_{\mu\nu}$ are the field strengths of
$B_{\mu}$ and $V_{\mu}$ respectively. All anomaly coefficients
depend on the coupling $g(m)$ at renormalization scale
$m$. The first term of (\ref{2.1}) is the internal trace
anomaly, where $\tilde{\beta}(g)$ is the numerator of the NSVZ beta
function \cite{nsvz}
$\tilde{\beta}(g(\mu))=-g^3\left[3T(G)-\sum_iT(R_i)
(1-\gamma_i(g(\mu)))\right]/16\pi^2 .$
Here $T(G)$ and $T(R_i)$ are the Dynkin indices of the adjoint
representation of $G$ and the representation $R_i$ of the chiral
superfield $\Phi^{\alpha}_{i}$, and $\gamma_i/2$ is the anomalous
dimension of $\Phi^{\alpha}_{i}$.

The various external anomalies are contained in the three
coefficients $\tilde{b}(g)$, $\tilde{c}(g)$ and $a(g)$. 
The free field
({\em i.e.} one-loop) values of $\tilde{c}$ and $a$ have been known  
for many years
\cite{duff}. 
In a free 
supersymmetric gauge theory with $N_v={\rm dim}\,G$ gauge multiplets 
and $N_\chi$ chiral
multiplets these values one has
$c_{UV}=(3N_V+N_\chi)/24$ and 
$a_{UV}=(9N_V+N_\chi)/48$.
If $T^j_i$ is the flavor matrix for the current $J_{\mu} (x)$
which is the $\bar{\theta} \theta$ component of the superfield
$\overline{\Phi}^i_{\alpha} T^j_i \Phi^{\alpha}_j$, and dim$R_i$
is the dimension of the representation $R_i$, 
the free-field value of $\tilde{b}$ is
$b_{UV} =\sum_{i,j} (\dim R_i) T^j_i T^i_j$.
The subscript $UV$ indicates that the free-field values are
reached in the ultraviolet limit of an asymptotically free
theory. 

The 
coefficients $\tilde{b}(g)$, $\tilde{c}(g)$ and $a(g)$
can be shown to be related to
the correlation functions 
$\langle J_{\mu} (x) J_{\nu} (0) \rangle$,
$\langle T_{\mu\nu} (x) T_{\rho\sigma}(0) \rangle$ 
and $\langle T_{\mu\nu} (x) T_{\rho\sigma}(y)
T_{\alpha\beta}(y)\rangle$
respectively \cite{AEF,Osban}.
Various positivity constarints \cite{AEF} for 
$\tilde{b}(g)$, $\tilde{c}(g)$ and $a(g)$
follow from this relation. 

In a supersymmetric theory in the external U$(1)$ gauge and
supergravity backgrounds discussed above, the divergence of the
$R^{\mu}$ current and the trace of the stress
tensor are components of a single superfield. Therefore the
supersymmetry partner of the trace anomaly $\Theta$ of
(\ref{2.1}) is
\begin{equation}
  \label{2.13}
  \partial_{\mu} (\sqrt{g} R^{\mu}) = - \frac{1}{3g^3}
  \widetilde{\beta} (g) (F \widetilde{F}) -
  \frac{\tilde{b}(g)}{48\pi^2} (B \widetilde{B}) +
  \frac{\tilde{c} (g) - a(g)}{24 \pi^2} R \widetilde{R} +
  \frac{5a(g) - 3 \tilde{c} (g)}{9 \pi^2} (V \widetilde{V})
\end{equation}
where $R$ and $\widetilde{R}$ on the right hand side are the
curvature tensor and its dual. The ratio $-2/3$ between the
first two terms of (\ref{2.1}) and (\ref{2.13}) is well known in
global supersymmetry, but the detailed relation of the anomaly
coefficients of the gravitational section was first derived in
\cite{AFG} by evaluating the appropriate components of the curved
superspace anomaly equation
\begin{equation}
  \label{2.14}
  \bar{D}^{\dot\alpha} J_{\alpha \dot{\alpha}} =
  \frac{1}{24\pi^2} 
\left(\tilde{c} W^2 - a \Xi \right)
\end{equation}
where $J_{\alpha \dot{\alpha}}$, $W^2$ and $\Xi$ are the
supercurrent, super-Weyl, and super-Euler superfields
respectively. This equation shows that all gravitational
anomalies are described by the two functions $\tilde{c}(g)$ and
$a(g)$, and this is also the reason why the coefficients of the third
and fifth terms of (\ref{2.1}) are related.

The infrared central charges
$b_{IR}$, $c_{IR}$ and $a_{IR}$ are determined as
the infrared limit (at $m\to 0$) of 
$\tilde{b}(g)$, $\tilde{c}(g)$ and $a(g)$.

These central charges are related to the
conventional U$(1)_R F^2$, U$(1)_R$ and U$(1)^3_R$
anomalies. 
It is useful to derive this relation using the
formalism of the all-orders anomaly-free $S^{\mu}$ current
\cite{kogan}. 
The external
 anomalies of this current can be clearly seen to agree in the
 infrared limit with those of the $R^{\mu}$ current which is in
 the same multiplet as the stress tensor, and thus part of the
 $N=1$ superconformal algebra of the infrared fixed point theory.

Gaugino fields are denoted by $\lambda^a (x) \, , \, a=1, \ldots
,\dim G$, and scalar and fermionic components of
$\Phi^{\alpha}_i(x)$ by $\phi^{\alpha}_i (x) $ and
$\psi^{\alpha}_{i} (x)$ respectively. The canonical $R^{\mu}$
current (which is the partner of the stress tensor), and the matter
Konishi currents $K^{\mu}_i$ for each representation are
\begin{eqnarray}
\label{2.15}
  R^{\mu} &=& \frac{1}{2} \overline{\lambda}^a \gamma^{\mu}
               \gamma^5 \lambda^a - \frac{1}{6} \sum_i
               \overline{\psi}^i_{\alpha} \gamma^{\mu} 
               \gamma^5 \psi^{\alpha}_{i} + \frac{2}{3}
               \sum_i \overline{\phi}^i_{\alpha} 
               \stackrel{\leftrightarrow}{D}_{\mu}
               \phi^{\alpha}_i \nonumber\\[1ex]
  K^{\mu}_i &=& \frac{1}{2} \sum_i \overline{\psi}_{\alpha}^i \gamma^{\mu}
                \gamma^5 \psi^{\alpha}_i +
                \sum_i \overline{\phi}^i_{\alpha}
                \stackrel{\leftrightarrow}{D}_{\mu} \phi^{\alpha}_i \, .
\end{eqnarray}
Conservation of both currents is spoiled by a classical
violation for any non-vanishing superpotential
$W$ and a chiral
anomaly.  
There is then a
(flavor singlet) all-order conserved current
($\partial_{\mu} S^{\mu} = 0$)
\begin{equation}
\label{2.20}
  S^{\mu} = R^{\mu} + \frac{1}{3} \sum_i (\gamma^{\ast}_i - \gamma_i) 
  K^{\mu}_i
 \end{equation}
where $\gamma^{\ast}_i$ stand for the infrared values of anomalous dimensions
of the chiral fields.

In physical correlators the infrared limit can be associated 
with large distance behavior.  
Therefore in the infrared (large distance) limit 
of correlators with an insertion of $R_{\mu}=
S_{\mu}-\frac{1}{3} \sum_i (\gamma^{\ast}_i - \gamma_i) 
K^{\mu}_i$ 
the contribution of the Konishi current decreases faster
than the contribution of the $S_{\mu}$ current
which has no anomalous dimension.
Thus the $S^{\mu}$ and $R^{\mu}$ operators and their
correlators agree in the long distance limit, as is required at
the superconformal $IR$ fixed point.
In the free $UV$ limit
$\gamma_i \to 0$
and the appropriate correlators can be computed 
perturbatively.

Because the current $S^{\mu}$ is exactly conserved without
internal anomalies, 't Hooft anomaly matching \cite{hooft} can be
applied to calculate the anomalies of its matrix elements with
other exactly conserved currents, such as $\partial_\mu
\langle S^\mu
T^{\rho\sigma} T^{\lambda\tau}\rangle $. 
By using the fact that $S$ and $R$ coincide at long
distances we have the chain of equalities
\begin{equation}
\label{2.26}
  \partial\langle RTT \rangle_{IR} = \partial\langle STT
  \rangle_{IR} = \partial\langle STT \rangle_{UV}
\end{equation}
where the last term simply includes the one loop graphs of the
current $S$ and gives the U(1)$_R$ anomaly coefficient quoted
in the literature. 
Similar arguments
justify the conventional calculation of of U$(1)_RFF$ and
U$(1)_R^3$ anomalies.

\section{Formulas for central charges}

The previous discussion enables us to write simple formulae for the quantities $b_{IR}$, $c_{IR}$ and $a_{IR}$
\begin{eqnarray}
b_{IR}&=&3\sum_{ij}
({\rm dim}\,R_i)(1-r_i) T_i^j\,T_j^i \label{abcIR} \nonumber \\  
c_{IR}&=&
\frac{1}{32}(4{\rm dim}\,G+\sum_i 
({\rm dim}\,R_i)(1-r_i)(5-9(1-r_i)^2) \nonumber \\
a_{IR}&=&
\frac{3}{32}(2{\rm dim}\,G+\sum_i 
({\rm dim}\,R_i)(1-r_i)(1-3(1-r_i)^2)). 
\label{generic}
\end{eqnarray}
The corresponding UV quantities are obtained from (\ref{abcIR}) by replacing
$r_i\rightarrow 2/3$.

In the presence of accidental symmetry the formulas for the
infrared values of
$a$, $b$ and $c$ have to be modified.
The appearance of accidental symmetry is associated with
decoupling \cite{accid} in the infrared
of a primary gauge invariant chiral composite field $M$.
In this case the infrared $R$ current can be determined as an
infrared limit of a linear combination
$R_{\mu}^{IR}=S_{\mu}+ A_{\mu}$, where $A_{\mu}=\lambda J^{(M)}_{\mu}$,
and $J^{(M)}_{\mu}$ is the current for the components of the 
superfield $M$.
The coefficient $\lambda$ is fixed by the condition that 
$R=2/3$ for the field $M$.
The corrected infrared values \cite{AEF} of the central charges are
\begin{eqnarray}
b_{IR}&=&b_{IR}^{(0)}+3
\,T^i_j T_i^j~\left(r_M-\frac{2}{3}\right) ,\cr
a_{IR}&=&a_{IR}^{(0)}+
\frac{\rm dim~ M}{96}(2-3r_M)^2
(5-3r_M),\cr
 c_{IR}&=&c_{IR}^{(0)}+
\frac{\rm dim~ M}{384}
(2-3r_M)[(7-6r_M)^2-17].
\label{accid}
\end{eqnarray}
Here we denoted by $b_{IR}^{(0)}$, 
$a_{IR}^{(0)}$ and $c_{IR}^{(0)}$ the 
expressions for $b$, $a$ and $c$ given by equations (\ref{generic}),
and $r_M$ stands for the $S$ charge of the chiral field $M$,
specifically the sum of the $S$ charges of its elementary constituents.
Here $T^i_j$ stands for the flavor generator associated with $b$.

In what follows we mainly focus on the positivity constraint 
$a_{UV} - a_{IR}>0$.
As explained above, the gravitational effective action depends on
the functions $a$ and $c$.
It is natural to assume that a candidate $C$-function
measuring the irreversibility of the RG flow
may be a universal model independent linear combination
$C=ua+vc$.
With 
our formulas for the infrared 
values of $c$ and $a$ it is easy to show
$a_{UV}-a_{IR} >0$ is the only universal $a$-theorem candidate,
so that $C=a$.

It is easy to check that in the absence of accidental symmetries
in the models with a unique non-anomalous 
$S_\mu$ current $a_{UV} - a_{IR} \geq 0$ if $r_i \leq 5/3$ for all $\Phi^i$.
Remarkably,
$r_i \leq 5/3$ for all renormalizable models, so the
$a$-theorem is always satisfied.
Most 
of the positivity conditions,
especially the $a$-theorem, can be verified essentially by
inspection of the tables of $R$-charges presented in the
literature on the various models. 
Actually, in
many cases one
can prove that $r_i<5/3$ as a consequence of asymptotic freedom
in absence of 
accidental symmetry (i.e. when all $r_i\geq 1/3$).

An interesting sutuation is a flow between two
non-trivial conformal theories.
A conformal fixed point is characterized
by the values of $b$,
$c$ and $a$.
One may consider the
flow $a_{UV} - a_{IR}$ for a theory which
interpolates between two
interacting conformal fixed points.
Such an interpolation may be obtained by deforming
a superconformal theory with a relevant operator
which generates an RG flow driving the theory to another
superconformal fixed point.
(Such a deformation may be, for example due to higgsing 
the gauge group.)
Since we know the conformal theories at both
ultraviolet and infrared limits of this
interpolating theory,
the computation simply requires subtraction of 
the end-point central charges.

In refs. [1,2] the positivity constraints 
(including $a_{UV} - a_{IR}>0$) 
have been checked for most of 
renormalizable models 
with uniquely determined  
$S_\mu$ current known to the authors by that time.
The list of models includes

$\bullet$ Models with one type of irreducible representation:
the SU$(N_c)$ series,
SO$(N_c)$ series \cite{seiberg},
Sp$(2N_c)$ series \cite{Intsei},
Pouliot Spin(7) model \cite{pouliot},
Distler-Karch models with exceptional groups \cite{distler},
Spin(7) Pouliot model with $N_f$ spinors {\bf 8},
$Q_i$, $G_2$ with $N_f$ {\bf 7},
$E_7$ Distler-Karch model: 4 fundamentals {\bf 56}, $Q_i$,
$E_6$ Distler-Karch model (I): 6 fundamentals {\bf 27},
$Q_i$, $E_6$ Distler-Karch model (II): 
$3\times ({\bf 27}+{\bf \overline{27}})$ fundamentals $Q_i$, 
$F_4$ Distler-Karch model: 5 fundamentals {\bf 26}, $Q_i$, 
$F_4$ Distler-Karch model: 4 fundamentals {\bf 26}, $Q_i$, 
Spin(8) Distler-Karch model: 
$4\times ({\bf 8}_v+{\bf 8}_c+{\bf 8}_s)$ fundamentals $Q$.

$\bullet$ Models with two types of irreps
with uniquely determined $S$ current.
This set of models includes
Kutasov-Schwimmer-type models for SU groups \cite{kut},
for SO and Sp 
gauge groups \cite{intr,leigh} ,
and the Pouliot Spin(7) model with $N_c+4$ flavors in $\bf 8$
and singlets \cite{pouliot}.

It is worth emphasizing that
our approach does not work for non-renoralizable models
because it is based on an interpolation between the IR critical theory 
and 
the UV {\it free} theory.

\section*{Acknowledgments}
I am very grateful to Damiano Anselmi, 
Josh Erlich, Dan Freedman and Marc Grisaru
for collaboration.

\end{document}